\documentclass[12pt, leqno]{article}
\usepackage[english]{babel}
\usepackage{amsmath}
\usepackage{amssymb}
\usepackage{amsfonts}
\usepackage{amsthm}
\usepackage[a4paper,left=2.4cm,right=2.4cm,top=2.7cm,bottom=2.7cm]{geometry}
\usepackage{amsfonts, amsmath, amssymb, amsthm, mathrsfs}
\usepackage{graphicx}
\usepackage{color}
\usepackage[sans]{dsfont}
\usepackage{amsmath,amssymb,bm}
\usepackage[T1]{fontenc}
\usepackage[english]{babel}
\usepackage{graphicx} 
\usepackage{booktabs} 
\usepackage{color}
\graphicspath{ {images/} }

\date{}

\title{A ``Garden of Forking Paths''  -- the Quantum Mechanics of Histories of Events \\
\vspace{1cm}
\small{dedicated to the memory of Raymond Stora\\
\vspace{0.5cm}
\textit{``I leave to several futures (not to all) my garden of forking paths'' \footnote{Jorge Luis Borges, \textit{``El jard\'{i}n de senderos que se bifurcan,''} Editorial Sur, 1941} }}}

\begin{document}

\maketitle
\begin{center}
\author{Philippe Blanchard $^{1}$, J\"{u}rg Fr\"{o}hlich $^{2}$, Baptiste Schubnel $^{3}$}
\end{center}

{$^{1}$ Faculty of Physics, University of Bielefeld, Universit\"{a}tsstr. 25, D-33615 Bielefeld}

 {$^{2}$ Theoretical Physics, ETH Zurich, Wolfgang Pauli Strasse 27, CH-8093 Zurich}
 
 {$^{3}$ SBB -- Swiss Federal Railways, Wylerpark, CH-3014 Bern}
 
\begin{abstract}
We present a short survey of a novel approach, called \textit{``ETH approach''}, to the quantum theory of events happening in isolated physical systems and to the effective time evolution of states of systems featuring events. In particular, we attempt to present a clear explanation of what is meant by an ``event'' in quantum mechanics and of the significance of this notion. We then outline a theory of direct (projective) and indirect observations or recordings of physical quantities and events. Some key ideas underlying our general theory are illustrated by studying a simple quantum-mechanical model of a mesoscopic system.
\end{abstract}

\tableofcontents

\section{Introduction -- walking out of the quantum maze}
\label{intro}

\subsection{Introductory remarks}
Recent years have seen enormous progress in setting up beautiful experiments that successfully test fundamental features of quantum mechanics. Furthermore, there have been substantial new developments in the areas of quantum information theory and its practical uses and of quantum computing. These advances have made renewed studies of the foundations of quantum theory commendable and, perhaps, even somewhat fashionable -- after a long period during which such studies were seen with suspicion. 

Unfortunately, the success of recent efforts to clarify the message and interpretation of quantum mechanics and to formulate this theory in a logically coherent way is rather limited. Much confusion and disorientation still surround its foundations, even among professional physicists -- so much so that many mathematicians do not want to think about it. There are many wrong or misleading prejudices. To mention one example, we tend to teach to our students that, in the Schr\"{o}dinger picture, the quantum-mechanical time evolution of states of physical systems is described by the Schr\"{o}dinger equation for a wave function (or the Liouville equation for a density matrix), and that the Schr\"{o}dinger picture and the Heisenberg picture are equivalent. Well, when stated in this generality and in case we wish to describe the time evolution of systems featuring \textit{events} (amenable to observation), \textit{nothing could be farther from the truth}; see subsect. \ref{ETH}!

Given that quantum mechanics was discovered ninety years ago, the present rather low level of understanding of its deeper meaning may be seen to represent some kind of intellectual scandal. We would like to help, in modest ways, to alleviate some of the confusion surrounding this most important theory. 

Fairly shortly before his death, our unforgettable mentor and friend \textit{Raymond Stora} developed a lively interest in questions concerning the foundations of quantum mechanics. We feel that it is fitting to dedicate a paper on this subject to his memory.

\subsection{Some fundamental questions and problems}
In our courses on quantum mechanics, we tend to describe physical systems, $S$, as pairs of a Hilbert space, $\mathcal{H}$, of pure state vectors, and a unitary propagator, $(U(t,s))_{t,s\in \mathbb{R}}$, describing the time evolution of states (from time $s$ to time $t$). Unfortunately, these data encode hardly any interesting invariant data about $S$, besides spectral properties of the unitary operators $U(t,s)$, which would enable one to draw conclusions about physical properties of $S$. Moreover, they give the erroneous impression that quantum mechanics might be a deterministic theory, because the Schr\"{o}dinger equation is a deterministic evolution equation. These observations give rise to the following

\vspace{1cm}
\textit{Fundamental questions and problems:}
\begin{itemize}
\item{What does one have to add to the data described above to arrive at a mathematical structure that -- through interpretation -- can be given physical meaning \textit{without the intervention of ``observers'' or addition of ad-hoc postulates concerning ``measurements'' to the theory?}}
\item{What is the origin of the \textit{intrinsic randomness} in quantum theory, given the deterministic character of the Schr\"{o}dinger equation? In which way does quantum randomness differ from classical randomness? }
\item{How are \textit{``states'', ``observables''} and \textit{``events''} defined in quantum mechanics; what is the meaning of these notions? Do we understand the \textit{time-evolution} of ``states'' and of ``observables'' of physical systems in quantum mechanics, and what does it have to do with solving the Schr\"{o}dinger equation?}
\item{What is meant by an \textit{``isolated system''} in quantum mechanics, and why is this an important notion? (\textit{Answer:} Because only for isolated systems we understand, in a general way, how to describe the \textit{time evolution of ``observables''}.) Is it understood, in quantum mechanics, how to prepare a system
$S$ in a prescribed initial state? Etc.}
\end{itemize}
Answers to these important questions, except to the last one, are sketched in the following sections. (A fairly detailed discussion of the last question can be found in \cite{FFS, FS1}.)

\subsection{Purpose of analysis}

Besides addressing the questions raised in the last subsection, the main purpose of this paper is to sketch or review arguments in favor of some of the following basic claims.
\begin{itemize}
\item{In quantum mechanics, the ``state'' of a system -- as conventionally defined -- does $not$ describe ``what is'' or ``will be''; it does not have an ontological status. Rather it is a mathematical device enabling us to make bets about the most likely events seen to happen in the future. (The ``ontology'' lies in \textit{time-ordered sequences of events}, sometimes called \textit{``histories''}, not in ``states''.)}
\item{The success of a quantum theory of ``events'' (that can be detected through observations or measurements) hinges on our ability to update the state of a system in time in accordance with events observed in the past, i.e., on a description of the time evolution of states in the presence of events, which observers can, in principle, record with the purpose to optimize their predictions of future events.}
\item{Our description of the time evolution of states of systems exhibiting events exploits a fundamental mechanism of \textit{``loss of access to information''} (for short, ``information loss'') and of \textit{entanglement with degrees of freedom carrying away ``inaccessible (lost) information''}. This mechanism allows for the evolution of pure states into mixtures.}
\item{There is no reason to expect that there are \textit{``information- or unitarity paradoxes''} in quantum mechanics. In fact, the quantum-mechanical time evolution of states of physical systems exhibiting information loss and featuring events  that can be recorded is \textit{never unitary}; (see subsect. \ref{ETH}).}
\item{Somewhat advanced mathematical concepts, such as functional analysis, in particular operator algebras (including type $III_{1}$- von Neumann algebras), functional integration, stochastic processes, elements of statistics, etc. have been invented to be \textit{used} in the study of Quantum Theory -- they do not represent a superfluous luxury.}
\end{itemize}
In the following sections, we sketch arguments in favor of some of these claims; (for a more detailed presentation we refer to \cite{FS2, BH, BFFS}). In particular, we outline a novel theory of events, observations and measurements in quantum mechanics based on two basic concepts: 
\begin{enumerate}
\item{Fundamental ``loss of access to information'' and entanglement with degrees of freedom that are no longer observable, i.e., carry away lost information.}
\item{Specification of a list of physical quantities characterizing possible events that can, in principle, happen, (depending on the state the system has been prepared in) and be recorded directly.}
\end{enumerate}

\section{Information loss and events in quantum mechanics}
\label{events}
We start this section with a somewhat pedestrian definition of physical systems in quantum mechanics, (subsect. 2.1). Afterwards, we introduce the concept of ``loss of access to information'', (subsect. 2.2). This will guide us towards a novel quantum-mechanical theory of events amenable to observation, (subsect. 2.3). Finally, in subsect. 2.4, we describe the time evolution of states in physical systems featuring events that can be recorded.

It should be emphasized that what we are trying to understand in this paper is \textit{Quantum Mechanics} -- pure and simple; we are \textit{not} trying to extend or generalize this theory.

\vspace{0.3cm}
\subsection{Definition of physical systems}
\label{Phys-syst}
\textit{Definition N-R:} In non-relativistic quantum mechanics, an isolated system, $S$, is characterized by the following data, (items 1 through 3):
\begin{itemize}
\item{1. A pair,
\begin{equation}
\label{1}
(\mathcal{H}, \lbrace U(t,s) \rbrace_{t,s \in \mathbb{R}}),
\end{equation}
of a Hilbert space $\mathcal{H}$ of pure state vectors and a unitary propagator $U$ with the usual properties:     $U(t,s)$ is a unitary operator on $\mathcal{H}$, for all pairs of times $(t,s)$, and 
$$U(t,t) = {\bf{1}}, \quad U(t,s)\cdot U(s,r) = U(t,r), \quad \forall t, s, r \text{   in   } \mathbb{R}.$$}
\item{2. A list,
\begin{equation}
\label{2}
\mathcal{O}_{S} = \lbrace \hat{X}_{i} \rbrace_{i \in I_{S}},
\end{equation}
of bounded self-adjoint operators $\hat{X}_{i}$ representing physical quantities of $S$  that could be recorded directly. We assume that $\mathcal{O}_{S}$ contains an identity element, ${\bf{1}}$, and that if $f$ is an arbitrary real-valued, bounded, continuous function on $\mathbb{R}$ and $\hat{X}$ is an arbitrary operator in $\mathcal{O}_{S}$ then $f(\hat{X})$ also belongs to $\mathcal{O}_{S}$.}
\end{itemize}

\textit{Remarks.} (i) In this paper, \textit{``physical quantities of a system $S$''} are always represented by self-adjoint (bounded) linear operators. \footnote{This is actually a feature common to all physical theories known to us -- quantum and classical.} If during a certain interval, $\mathcal{I}$, of time it is possible to unambiguously assign an objective value to a physical quantity of $S$ represented by an operator $\hat{X} \in \mathcal{O}_{S}$ we say that, during the time interval $\mathcal{I}$, an ``event'' is happening; namely the event that $\hat{X}$ has an objective value that could, in principle, be observed directly. What this means mathematically will be explained below.\\

(ii) Note that, in general, $\mathcal{O}_{S}$ is not an algebra; it is not even a linear space! Typically, 
$\mathcal{O}_{S}$ may be generated by just a few (possibly only finitely many) operators. 
Let $\mathcal{A}_{S}$ denote the algebra generated by $\mathcal{O}_{S}$ (closed in a $C^{*}$- norm). In simple examples of physical systems (see Eq. \eqref{X} and subsect. \ref{Kraus-meas} for a concrete model system), the operators in $\mathcal{O}_{S}$ all commute among themselves. We can then identify $\mathcal{A}_{S}$ with $\mathcal{O}_{S}$; and it is a well known theorem due to I. M. Gel'fand that, under this assumption,
\begin{equation}
\mathcal{O}_{S} \simeq \lbrace \text{functions on a compact Hausdorff space  } \mathcal{X}_{S} \rbrace
\end{equation}
The topological space $\mathcal{X}_{S}$ is called the \textit{spectrum} of $\mathcal{O}_{S}$.

Given an algebra $\mathcal{A}$ of operators, a maximal abelian subalgebra of $\mathcal{A}$ is a commutative subalgebra, $\mathcal{M} \subseteq \mathcal{A}$, with the property that the subalgebra of operators in $\mathcal{A}$ that commute with \textit{all} operators in $\mathcal{M}$ is equal to $\mathcal{M}$.
In order to keep this paper reasonably short and easy to read, we introduce the following\\

\vspace{-0.3cm}
\underline{\textit{Simplifying Assumption:}}\\
Every maximal abelian algebra, $\mathcal{M}$, contained in 
$\mathcal{A}_{S}$ is generated by a \textit{finite} family of commuting orthogonal projections, 
$\lbrace \Pi_{\xi_1},..., \Pi_{\xi_N} \rbrace$. Then $\mathcal{M} = C(\mathcal{X})$, where 
$\mathcal{X}=\lbrace \xi_1,...,\xi_N \rbrace$ is the spectrum of $\mathcal{M}$. It is assumed that there is \textit{at least one} maximal abelian subalgebra, $\mathcal{M}_{S}$, in $\mathcal{A}_{S}$ with the property that \textit{all} self-adjoint elements of 
$\mathcal{M}_{S}$ belong to $\mathcal{O}_{S}$. However, there may be several such maximal abelian algebras, $\lbrace \mathcal{M}^{(i)}_{S}\rbrace_{i \in I_S}$, not commuting with each other. \footnote{Example: $\mathcal{M}_{S}^{(1)}=\lbrace \text{all bounded continuous functions of the position of a particle},  P \rbrace$, and $\mathcal{M}_{S}^{(2)}=\lbrace \text{all bounded continuous functions of the momentum of } P \rbrace$.}
 The orthogonal projections contained in an algebra 
$\mathcal{M}^{(i)}_{S}, i \in I_S,$ are called \textit{``possible events''}; any real linear combination of the orthogonal projections generating $\mathcal{M}^{(i)}_{S}$ is then a physical quantity, $\hat{X}_i$, belonging to $\mathcal{O}_{S}$. \footnote{A more general analysis of the role of maximal abelian subalgebras of 
$\mathcal{A}_{S}$ in our formulation of quantum theory, \textit{not} assuming that they are generated by finitely many projections, will be presented elsewhere.}\\

(iii) The occurrence of events in a system $S$ does \textit{not} depend on the presence of ``observers''; i.e., our formulation of quantum mechanics does \textit{not} invoke ``observers'' who decide to measure some quantity (and may then disagree on exactly which quantity they would like to measure and when). But, of course, any useful physical theory must talk about objects and phenomena that intelligent beings can observe if they choose to do so, and it should help them to cope with the challenges of a changing world by enabling them to agree among themselves whether some events have happened and to make useful and plausible predictions about future events. -- That much about ``physical quantities'' (``observables''), ``(possible) events'', and philosophy!\\

\begin{itemize}
\item{3. At every time $t$, there exists a representation 
\begin{equation*}
\mathcal{A}_{S} \owns \hat{X} \mapsto X(t)
\end{equation*}
of the algebra $\mathcal{A}_{S}$ by operators, $X(t)$, acting on the Hilbert space $\mathcal{H}$ with the property that $\hat{X}^{*}$ is represented by the operator $X(t)^{*}$; in particular, if $\hat{X}$ is self-adjoint then $X(t)$ is a self-adjoint operator on $\mathcal{H}$. The operators $X(t)$ and 
 $X(s)$ are unitarily conjugated to each other by the propagator of $S$, i.e.,  
 \begin{equation*}
 X(t)=U(s,t)X(s)U(t,s), \quad \text{  for times  } s,t \in \mathbb{R}, \text{   }\hat{X} \in \mathcal{A}_{S}.
\end{equation*}
}
\end{itemize}

By $\mathcal{A}_{S}(t)$ we denote the algebra $\lbrace X(t) \vert \hat{X} \in \mathcal{A}_{S} \rbrace \subseteq B(\mathcal{H})$, where, as usual, $B(\mathcal{H})$ denotes the \textit{algebra of all bounded operators on the Hilbert space} $\mathcal{H}$.\\

Possible events observable at times $\geq t$ generate an algebra $\mathcal{E}_{\geq t}$:
\begin{equation}
\label{eventalg}
\mathcal{E}_{\geq t} := \lbrace \text{linear combinations of  } \prod_{i} X_{i}(t_i)\vert \hat{X}_{i} \in \mathcal{O}_{S}, t_{i} \geq t \rbrace ^{-},
\end{equation}
with $\mathcal{E}:=\overline{\mathcal{E}_{>-\infty}}$. For concreteness, we assume that the closure is taken in the weak operator topology on $B(\mathcal{H})$. \footnote{A sequence, or net,
$(A_i)_{i\in I}$ of bounded operators on $\mathcal{H}$ is said to converge weakly iff  
$(\langle \psi, A_{i} \varphi \rangle)_{i \in I}$ converges, for arbitrary vectors $\psi$ and $\varphi$ in 
$\mathcal{H}$. The algebras $\mathcal{E}_{\geq t}$ and $\mathcal{E}$ are von Neumann algebras, because they are closed under weak convergence. In the following, it is convenient to work with von Neumann algebras. But the reader is kindly asked not to worry about this technicality.}
\\

With a view towards an extension of our formalism to \textit{relativistic quantum (field) theory}, we briefly outline a somewhat more general notion of physical systems.

\textit{Definition R:} In quantum theory, a general isolated physical system $S$ is characterized by the following data:
\begin{itemize}
\item{1. A list,
\begin{equation*}
\mathcal{O}_{S} = \lbrace \hat{X}_{i} \rbrace_{i \in I_{S}},
\end{equation*}
of bounded self-adjoint operators $\hat{X}_{i}$ representing physical quantities of $S$. As before, we let 
$\mathcal{A}_{S}$ denote the $C^{*}$- algebra generated by $\mathcal{O}_{S}$, and we continue to impose the simplifying assumption formulated in Remark (ii) after item 2, above, etc.}
\item{2. A net $( \mathcal{E}_{\mathcal{I}})_{\mathcal{I} \subset \mathbb{R}}$ of (von Neumann) algebras, 
$\mathcal{E}_{\mathcal{I}}$, indexed by time intervals $\mathcal{I}$, with the interpretation that 
$\mathcal{E}_{\mathcal{I}}$ is generated by possible events localized in the time interval $\mathcal{I}$. 
This net is assumed to have the property that if $\mathcal{I} \subset \mathcal{I}'$ then 
$\mathcal{E}_{\mathcal{I}} \subset \mathcal{E}_{\mathcal{I}'}$. We define 
\begin{equation}
\label{eventalg}
\mathcal{E}_{\geq t} := \overline{\bigvee_{\mathcal{I}\subseteq [t, \infty)} \mathcal{E}_{\mathcal{I}} }, \qquad \mathcal{E}:= \overline{\mathcal{E}_{> -\infty}}.
\end{equation}
In \eqref{eventalg}, the closure is taken in the weak operator topology on $B(\mathcal{H})$.} 
\item{3. For every time $t\in \mathbb{R}$ there is a $^{*}$representation
\begin{equation}
\label{reps}
\mathcal{A}_{S} \owns \hat{X} \mapsto X(t) \in \mathcal{E}_{\geq t}
\end{equation}
of the algebra $\mathcal{A}_{S}$ by operators in $\mathcal{E}_{\geq t}$.} The representations of 
$\mathcal{A}_{S}$ corresponding to different times are unitarily equivalent.\\
 It is assumed, furthermore, that, for every $\hat{X} \in \mathcal{O}_{S}$ and every $\varepsilon > 0$, there exist a finite duration $\tau = \tau(\hat{X},\varepsilon)< \infty$ and an operator 
 $X_{\varepsilon}(t) \in \mathcal{E}_{[t,t+\tau]}$ such that 
$$\Vert X(t) - X_{\varepsilon}(t) \Vert < \varepsilon.$$
\end{itemize}

\textit{Remark:} In Definition $R$, ``time'' refers to the proper time of an observer, and the net 
$\lbrace \mathcal{E}_{\mathcal{I}} \rbrace_{\mathcal{I} \subset \mathbb{R}}$ depends on the worldline of that observer; see Fig. 1, below. This does \textit{not} mean that the theory becomes ``observer-dependent''. But it does mean that one has to find out how one and the same sequence of events is seen by different observers, i.e., how to map the data concerning a sequence of events recorded by one observer to the data recorded by another observer. Luckily, for the purposes of the analysis presented here we do not need to address this problem, which lies somewhat beyond the scope of this paper.

The analysis presented in the following sections is based on \textit{Definition R}; (but no attempt is made to
present an analysis that takes into account the laws of relativity theory).

\subsection{Information loss}
\label{iloss}
The idea of ``information loss'' or, more precisely, ``loss of access to information'' is encapsulated in the following general assumption concerning the algebras $\mathcal{E}_{\geq t}, t \in \mathbb{R}$:
\begin{equation}
\label{infoloss}
B(\mathcal{H}) \supseteq \mathcal{E} \supseteq \mathcal{E}_{\geq t} \underset{\not=}{\supset} \mathcal{E}_{\geq s} \supseteq \mathcal{A}_{S}(s), \qquad s>t.
\end{equation}
\hspace{5.9cm}{\textit{Information Loss!}\\

A precise formulation of ``Information Loss'' is to assume that if $s>t$ then
$$\hspace{4cm} \mathcal{E}_{\geq s}' \cap \mathcal{E}_{\geq t} \not= \emptyset, \hspace{4cm} (^{*})$$
where, for an algebra $\mathcal{A}$ of bounded operators acting on $\mathcal{H}$, $\mathcal{A}'$ is the algebra of all bounded operators on $\mathcal{H}$ commuting with all operators in $\mathcal{A}$. In fact, one expects that $\mathcal{E}_{\geq s}' \cap \mathcal{E}_{\geq t}$ is typically an infinite-dimensional algebra (at least for some $s>t$), an expectation extracted from the analysis of examples; see \cite{BRob, Schubnel}.\\ 
Property $(^{*})$ is far from obvious and appears to only hold in theories of systems with infinitely many degrees of freedom including \textit{massless ones}, such as photons or phonons. D. Buchholz and the late J. E. Roberts have presented a deep analysis of Property $(^{*})$ in quantum electrodynamics, formulated in the framework of algebraic quantum field theory; see \cite{BRob}. In their work, the analogue of the algebra 
$\mathcal{E}_{\geq t}$ is played by an algebra of bounded functions of the electromagnetic field smeared out with test functions with support in the forward light cone $\overline{V}^{+}_{P_{t}}$ erected over a space-time point $P_{t}$ at proper time $t$ that belongs to the worldline of an observer. They show that Property $(^{*})$ follows from Huyghens' Principle for the electromagnetic field and the existence of asymptotic electromagnetic field operators; see Fig. 1.\\

\hspace{1.5cm}
\includegraphics[width=7cm, height=5cm]{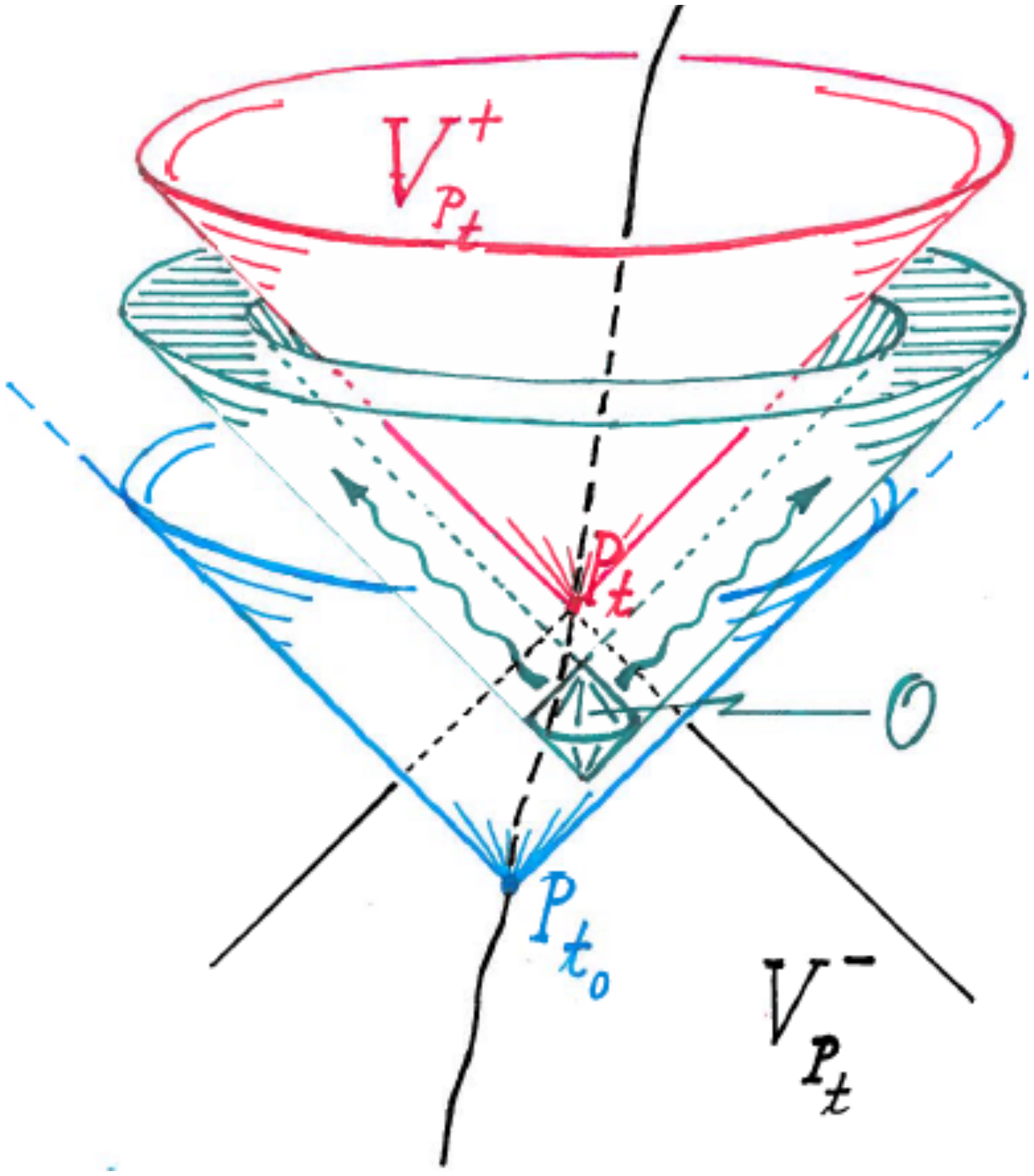}\\
\begin{center}
Figure 1: An illustration of Property $(^{*})$\\
\end{center}
Figure 1 indicates that $\mathcal{E}_{\geq t_0}$ properly contains $\mathcal{E}_{\geq t}$, for $t>t_0$, 
and that, asymptotically, flashes of light emitted from region $\mathcal{O}$ belong to 
$\mathcal{E}_{\geq t}' \cap \mathcal{E}_{\geq t_0}$.

Information Loss, in the sense of Eq. \eqref{infoloss} (with Property ($^{*}$) valid for \textit{some} $s>t$), holds in many models of \textit{non-autonomous} systems describing a small system (e.g., an $n$-level atom) alternatingly coupled to various mutually independent dispersive media (e.g., the quantized electromagnetic field, or the phonons of a dynamical crystal lattice) during finite intervals of time; see \cite{Schubnel}. Here we briefly sketch the example of a mesoscopic system consisting of a T-shaped conducting wire ending in three reservoirs denoted by $D_L$, $D_R$ and \mbox{``$e^{-}$ gun''}; see Figure 2.

\hspace{2cm}
\includegraphics[width=8.5cm, height=6.5cm]{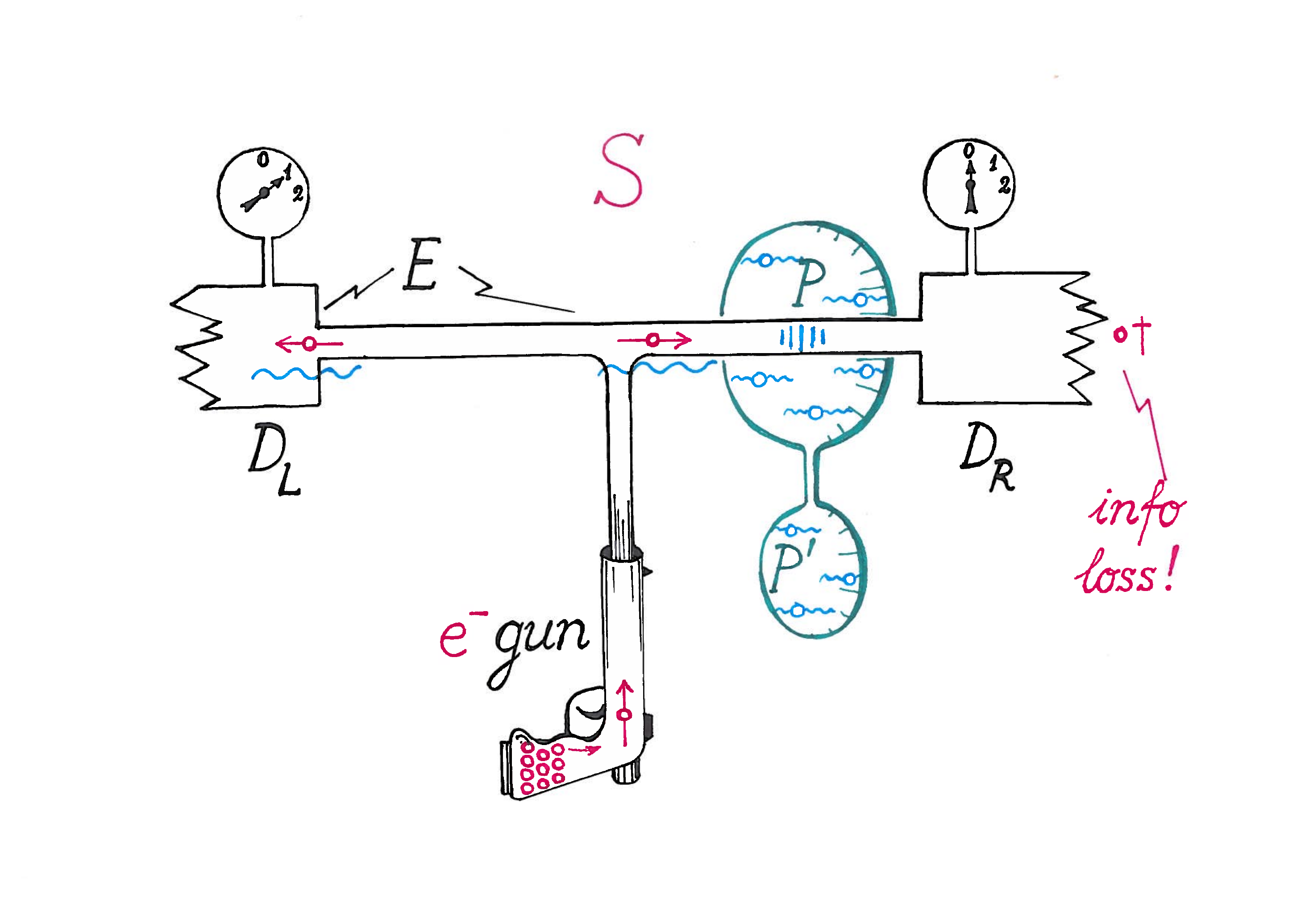}
\begin{center}
Figure 2: A mesoscopic system
\end{center}

The reservoir ``$e^{-}$ gun'' has a higher chemical potential than the reservoirs $D_L$ and $D_R$. Hence ``$e^{-}$ gun'' emits electrons at a certain rate that move through the T-shaped wire until they dive into one of the reservoirs $D_L$ or $D_R$ where they disappear for ever. Before they disappear they trigger detectors that emit a signal (flash of light or sound wave) whenever an electron has arrived at $D_L$ or $D_R$, respectively. In this example, the system $S$ is the composition of the equipment $E$ with a quantum dot $P\vee P'$ in a semi-conductor matrix. The electric charge localized inside the component $P$ of the quantum dot, which can fluctuate by electron exchange between the components $P$ and $P'$, determines the a-priori probability by which an electron traveling through the T-shaped wire will dive into $D_R$. The equipment 
$E$ consists of the three reservoirs, ``$e^{-}$ gun'', 
$D_L$ and $D_R$, the T-shaped wire, and the detectors at the entrance gates to $D_L$ and $D_R$. The \textit{only} physical quantity of $S$ that can be observed directly is the flash of light or sound emitted by the detector on the left or the right whenever an electron dives into $D_L$ or $D_R$, respectively. Mathematically, this quantity can be represented by the operator 
\begin{equation}
\label{X}
 \hat{X}= {\bf{1}}_{P\vee P'} \otimes \left( \begin{array}{cc}{\bf{1}} & 0\\
                                                                                                       0 & {\bf{-1}}
                                                                                                       \end{array} \right),
\end{equation}
which has the (infinitely degenerate) eigenvalues $\xi=\pm 1$, with 
\begin{center}
$\xi=+1 \leftrightarrow D_{L}$ clicks, \qquad $\xi=-1 \leftrightarrow D_{R}$ clicks.
\end{center}
The family $\mathcal{O}_{S}$ of operators consists of all bounded functions of the operator $\hat{X}$; 
its spectrum, $\mathcal{X}_{S}$, consists of two points, $\lbrace -1,+1 \rbrace$. Access to the ``information'' represented by an electron that travels through the T-shaped wire is \textit{lost}, as soon as that electron has dived into one of the reservoirs $D_L$ or $D_R$. (To make this precise one must assume that the detectors have infinitely many degrees of freedom.)

This example is discussed in much detail in \cite{BFFS}. It illustrates how properties of a physical system 
$S$ -- in the example, the charge of the dot $P$ -- can be determined \textit{indirectly} through a long sequence of repeated observations of physical quantities represented by operators in $\mathcal{O}_{S}$. Results from \cite{BFFS} concerning this example are summarized in section 3.
(Our efforts have been stimulated by the experiments described in \cite{Haroche}; see also \cite{BB}.)

\subsection{Direct detection of events -- projective recordings of physical quantities}
\label{events}
Let $\lbrace \mathcal{M}_{S}^{(i)} \rbrace_{i \in I_{S}}$ denote those maximal abelian subalgebras of 
$\mathcal{A}_{S}$ that belong to $\mathcal{O}_{S}$. In this subsection, we clarify what it means that a physical quantity of a system $S$ represented by an operator $\hat{X} \in \mathcal{M}_{S}^{(i)}$, for some $i \in I_S$, is recorded or measured directly (or ``projectively'') around some time $t$, i.e., that $\hat{X}$ has an objective value at or around time $t$. We will explain the roles played by ``information loss'', in the sense of Eq. \eqref{infoloss}, and of entanglement of observable degrees of freedom of $S$ with inaccessible (``lost'') degrees of freedom.

Let $ \xi_1< \cdots < \xi_N$ denote the eigenvalues of the operator
$\hat{X}$, and let $\Pi_{\xi_1}, ..., \Pi_{\xi_N}$ be the corresponding spectral projections, with
$\Pi_{\xi_j} \in \mathcal{M}_{S}^{(i)}, \forall j=1,...,N$. These projections have the interpretation of  ``possible events''; ($\Pi_{\xi_i}$ corresponds to the possible event that the physical quantity represented by the operator $\hat{X}$ is observed, at some time $t$, to have the value $\xi_j$). For the mesoscopic system considered in subsection \ref{iloss} (see Figure 2), $\mathcal{O}_{S} \equiv \mathcal{M}_{S}$ is generated by a single operator, $\hat{X}$, with only two eigenvalues $\xi= \pm 1$.

Let $X(t) \in \mathcal{E}_{\geq t}$ be the operator on 
$\mathcal{H}$ representing $\hat{X}$. Then
\begin{equation}
\label{spect-thm}
X(t) = \sum_{j=1}^{N} \xi_{j} \text{   } \Pi_{\xi_j}(t),
\end{equation}
where $\Pi_{\xi_j}(t)$ is the spectral projection of $X(t)$ corresponding to the eigenvalue 
$\xi_{j}$; (the eigenvalue $\xi_{j}$ is independent of time $t$, while the projections $\Pi_{\xi_j}(t)$ representing the projection $\Pi_{\xi_j} \in \mathcal{M}_{S}^{(i)}$ depend on $t$, but are all unitarily conjugate to one another, for every $j=1,...,N$).\\
It is compatible with the ``Copenhagen interpretation'' of quantum mechanics (whatever this interpretation may be, in more precise terms) to say that if the physical quantity corresponding to the operator $\hat{X} \in \mathcal{M}_{S}^{(i)}$ has an objective value around some time $t$, then the \textit{state} of $S$,
\begin{equation}
\label{state}
\rho(\cdot) = Tr_{\mathcal{H}}(P\cdot), \text{    where   } P \text{    is a density matrix on   } B(\mathcal{H}),
\end{equation}
\textit{when restricted to the algebra} $\mathcal{E}_{\geq t}$, is indistinguishable from an \textit{incoherent superposition} of eigenstates of the operator $X(t)$, in the following precise sense: Let 
$$\rho_{t} := \rho|_{\mathcal{E}_{\geq t}},$$ then
\begin{equation}
\label{mixtstate}
\rho_{t}(A)= \sum_{j=1}^{N} \rho_{t}(\Pi_{\xi_j}(t)A\Pi_{\xi_j}(t)) +O(\delta \Vert A \Vert), \qquad \forall A \in \mathcal{E}_{\geq t},
\end{equation}
for some constant $\delta$, with 
$$\delta \ll \underset{1 \leq i < j \leq N}{\text{min}}\vert \rho_{t}\big(\Pi_{\xi_i}(t)- \Pi_{\xi_j}(t)\big) \vert.$$
Eq. \eqref{infoloss} and entanglement with inaccessible degrees of freedom 
imply that the state $\rho_{t}$ is, in general, a \textit{mixed} state on $\mathcal{E}_{\geq t}$, \textit{even} if the state $\rho$ may be a \textit{pure} state on $B(\mathcal{H})$, so that Eq. \eqref{mixtstate} is by no means inconsistent. 

Given a state $\varphi$ on a von Neumann algebra $\mathcal{M}$, we define the \textit{centralizer}
(or stabilizer), $\mathcal{C}_{\varphi}$, of $\varphi$ to be the subalgebra of $\mathcal{M}$ defined by
\begin{equation}
\label{defcentral}
\mathcal{C}_{\varphi}:= \lbrace A \in \mathcal{M} \vert \text{ad}_{A}(\varphi) =0 \rbrace,
\end{equation}
where
$$ \text{ad}_{A}(\varphi)(B):= \varphi([A,B]), \quad \text{   for arbitrary  } B \in \mathcal{M};$$
see the Appendix for further details. For $\mathcal{M}=\mathcal{E}_{\geq t}$ and $\varphi = \rho_{t}$, the centralizer is henceforth denoted by 
$\mathcal{C}_{\rho_t}$.\\
Let us assume that either the algebra $\mathcal{M}$ is isomorphic to a direct sum 
$\bigoplus_{i} B(\mathcal{H}_i)$, where $\mathcal{H}_i, i=1,2,...,$ are Hilbert spaces, (i.e., that $\mathcal{M}$ is of type $I$), or that $\varphi$ is a separating state on $\mathcal{M}$ (meaning that
$\varphi(A^{*}A) = 0$, for some $A\in \mathcal{M}$, implies that $A=0$). \footnote{If the state $\varphi$ is separating on the von Neumann algebra $\mathcal{M}$ then $\mathcal{C}_{\varphi}$ is seen to be the subalgebra of operators in $\mathcal{M}$ invariant under the modular automorphism group, $(\sigma^{\varphi}_{t})_{t \in \mathbb{R}}$, corresponding to $(\mathcal{M}, \varphi)$ \cite{BR}; see, e.g.,  \cite{Combes, FS2} and references given there.} Then there exists a linear map 
$E_{\varphi}: \mathcal{M} \rightarrow \mathcal{C}_{\varphi}$, called a \textit{conditional expectation} from 
$\mathcal{M}$ to $\mathcal{C}_{\varphi}$, with the following properties:
$$ E_{\varphi}(A^{*}A) \geq 0, \quad \forall A \in \mathcal{M},$$
$$ E_{\varphi}(XAY) = X E_{\varphi}(A) Y, \quad \forall X, Y \in \mathcal{C}_{\varphi}, \forall A \in \mathcal{M},
$$
$$E_{\varphi}(A)^{*} E_{\varphi}(A) \leq E_{\varphi}(A^{*}A), \quad \forall A \in \mathcal{M}.$$
Let $\mathcal{Z}_{\varphi}$ denote the \textit{center} of $\mathcal{C}_{\varphi}$, (i.e., the algebra of all operators in $\mathcal{C}_{\varphi}$ commuting with all operators in $\mathcal{C}_{\varphi}$). Under the same assumptions, there also exists a conditional expectation $e_{\varphi}$ from $\mathcal{M}$ to 
$\mathcal{Z}_{\varphi}$ with the same properties as those of $E_{\varphi}$. 
(The general theory of conditional expectations in von Neumann algebras is developed in \cite{Dixmier, Combes, Takesaki}; applications to the centralizer of a von Neumann algebra can be found in \cite{Herman, Haagerup} and in references quoted therein.)
The conditional expectations from $\mathcal{E}_{\geq t}$ to $\mathcal{C}_{\rho_t}$ and from 
$\mathcal{E}_{\geq t}$ to $\mathcal{Z}_{\rho_t}$, the center of $\mathcal{C}_{\rho_t}$, are denoted by $E_{\rho_t}$ and $e_{\rho_t}$, respectively.

Let $X(t), \xi_j \text{   and   }\Pi_{\xi_j}(t), j=1,...,N,$ be as in Eq. \eqref{spect-thm}. It is not hard to show that
\begin{equation}
\label{centralizer}
\text{Eq. \eqref{mixtstate}} \Leftarrow \Vert E_{\rho_t}(\Pi_{\xi_j}(t)) - \Pi_{\xi_j}(t) \Vert \leq \delta', \quad \forall j=1,...,N, \end{equation}
for some $\delta' = O(\delta/N)$. This and the next claim are proven in the Appendix.

Obviously, Eq. \eqref{mixtstate} also holds if
\begin{equation}
\label{center}
\Vert e_{\rho_t}(\Pi_{\xi_j}(t))- \Pi_{\xi_j}(t) \Vert \leq \delta', \quad \forall j=1,...,N.
\end{equation}

We are now prepared to formulate the

\vspace{0.3cm}
{{\bf{\textit{Fundamental axiom of events in quantum mechanics:}}}

\vspace{0.2cm}
Let $\mathcal{P}:=\lbrace \Pi_{\xi_1},...,\Pi_{\xi_N} \rbrace$ be a partition of unity in $\mathcal{A}_{S}$, (i.e.,  
$\sum_{j=1}^{N} \Pi_{\xi_j} = {\bf{1}}\vert_{\mathcal{A}_{S}}$) consisting of commuting orthogonal projections that are contained in some maximal abelian subalgebra $\mathcal{M}_{S}^{(i)} \subseteq \mathcal{O}_{S}, i \in I_S$. These projections have the physical interpretation of \textit{``possible events''}, and any real linear combination of them is an operator, $\hat{X}\in \mathcal{O}_{S}$, representing a \textit{physical quantity} of $S$. Given a state $\rho$ (on the algebra $\mathcal{E}$) which the system $S$ has been prepared in, we propose to define what it means that \textit{one} out of these $N$ possible events actually happens (or materializes) around some later time $t$.\\
 We fix a ``threshold, $\Delta_{t}$, for detection (of an event) at time $t$'' satisfying
\begin{equation}
\label{threshold}
0< \Delta_{t} \ll \underset{i\not=j=1,...,N}{\text{min}}\text{   } \vert \rho_{t}(\Pi_{\xi_i}(t) - \Pi_{\xi_j}(t)) \vert,
\end{equation}
where $\rho_{t}=\rho\vert_{\mathcal{E}_{\geq t}}$, and $\Pi_{\xi_j}(t) \in \mathcal{A}_{S}(t)$ is the orthogonal projection on the Hilbert space $\mathcal{H}$ representing the projection $\Pi_{\xi_j}\in \mathcal{P}$. Let $\mathcal{P}(t):=\lbrace \Pi_{\xi_1}(t),..., \Pi_{\xi_N}(t) \rbrace$.\\
The fundamental axiom has two parts:
\begin{itemize}

\item{1. \textit{Occurrence of Events in Quantum Mechanics:}\\
\textit{One} of the (possible) events $\Pi_{\xi_1},....,\Pi_{\xi_N}$ happens (materializes) around time $t$ -- put differently, the physical quantity $\hat{X} = \sum_{j} \xi_{j} \Pi_{\xi_j}$ has an \textit{objective value} around time $t$ -- iff 
\begin{equation}
\label{distance}
\text{dist}(\mathcal{P}(t), \mathcal{Z}_{\rho_t}):= \underset{j=1,...,N}{\text{max}} \Vert e_{\rho_t}(\Pi_{\xi_j}(t))-\Pi_{\xi_j}(t) \Vert \leq \Delta_{t}/N.
\end{equation}}
\end{itemize}

\textit{Remarks.} (i) \textit{Time of occurrence of events:} Obviously, Eq. \eqref{distance} implies Eq. \eqref{mixtstate}.
The earliest time when a possible event in $\mathcal{P}$ can materialize is the smallest time $t=t_{\text{min}}$ at which inequality \eqref{distance} holds, after the preparation of $S$ in state $\rho$. Let $\mathcal{I}_{t_{\text{min}}}$ be the largest interval of time containing $t_{\text{min}}$ such that inequality \eqref{distance} holds for all 
$t \in \mathcal{I}_{t_{\text{min}}}$. Then one of the possible events $\lbrace \Pi_{\xi_j} \rbrace_{j=1}^{N}$ happens in $\mathcal{I}_{t_{\text{min}}}$. Most likely it happens around the time, $t_{*}$, minimizing 
the function $\text{dist}(\mathcal{P}(t), \mathcal{Z}_{\rho_t})$ defined in Eq. \eqref{distance}.

(ii) \textit{Duration of events:} Let $\tau = \tau(\mathcal{P}(t_{*}))$ be such that there are self-adjoint operators 
$$\lbrace \Pi_{j}(t_{*}, \tau) \rbrace_{j=1}^{N} \subset \mathcal{E}_{[t_{*}, t_{*}+\tau]}, \quad \text{    with}\quad
\Vert \Pi_{\xi_j}(t_{*}) - \Pi_{j}(t_{*}, \tau) \Vert \leq \Delta_{t_{*}}/N,$$
see item 3 of \textit{Definition R}, subsect. 2.1.
Then the \textit{duration of the event} happening around time $t_{*}$ is given by $\tau$.

(iii) \textit{A simple special case:} If the algebra $\mathcal{E}_{\geq t}$ is of type $I$ (which, alas, it usually won't be!) then the state $\rho_t$ can be represented by a density matrix, $P_t \in \mathcal{E}_{\geq t}$. Let
$$P_{t}=\sum_{j=1}^{N} p_j(t) \pi_{j}(t),$$
be the spectral decomposition of $P_t$, where the operators $\pi_{j}(t)$ are the spectral projections of
 $P_t$, and
$$\quad  p_{N}(t)>...>p_{1}(t)\geq0, \quad \text{   with   }\quad \sum_{j=1}^{N} p_{j}(t) \text{  dim  }(\pi_{j}(t)) =1.$$
Then \textit{one} of the possible events $\Pi_{\xi_1},...,\Pi_{\xi_N}$ happens around time $t$ iff
$$\underset{j=1,...,N}{\text{max}}\Vert \Pi_{\xi_j}(t) - \pi_{j}(t) \Vert \leq \Delta_{t}/N, \quad \text{with }\quad 
\Delta_{t} < \underset{j=2,...,N}{\text{min}} \big(p_{j}(t) - p_{j-1}(t)\big).$$

\begin{itemize}
\item{2. \textit{Randomness in Quantum Mechanics:}\\
Under the condition that \eqref{distance} holds at some time $t=t_{*} \in \mathcal{I}_{t_{\text{min}}}$, the probability that the possible 
event $\Pi_{\xi} \equiv \Pi_{\xi_j} \in \mathcal{P}$ actually materializes at time $t_{*}$ is given by 
\begin{equation}
\label{Born}
p_{\xi}(t_{*})= \rho(\Pi_{\xi}(t_{*}))
\end{equation}
\begin{center}
\textit{Born's Rule}
\end{center}
If the event corresponding to the projection $\Pi_{\xi} \in \mathcal{P}$ is detected to have happened at time $t_{*}$ then the state 
\begin{equation}
\label{collapse}
\rho_{\xi,t_{*}}(\cdot) := p_{\xi}(t_{*})^{-1} \text{   } \rho_{t_{*}}\big(\Pi_{\xi}(t_{*}) \cdot \Pi_{\xi}(t_{*}) \big)
\end{equation}
must be used for improved predictions of future events at times $>t_{*}$; i.e., the state of $S$ on the algebra, 
$\mathcal{E}_{\geq t_{*}}$, of possible events after time $t_{*}$, \textit{conditioned on the event corresponding to $\Pi_{\xi}$ to have materialized at time $t_{*}$}, is given by $\rho_{\xi,t_{*}}$.
\begin{center}
\textit{``Projection-, or Collapse Postulate''}
\end{center}}
\end{itemize}
\vspace{0.3cm}

\textit{Remarks, ctd.:} (iv) Apparently, if it is known that an isolated system $S$ was prepared in a state $\rho$ before the earliest event has happened, then the quantum theory of $S$ \textit{predicts} at or around what time $t_{*}$ the first event will occur, for what duration, $\tau$, the event will last, and to \textit{which family, 
$\mathcal{P}$, of possible events} that event belongs to. (We recall that $\mathcal{P}$ is contained in a maximal abelian subalgebra $\mathcal{M}_{S}^{(i)} \subseteq \mathcal{O}_{S}$ of $\mathcal{A}_{S}$.) But \textit{which event} from the family $\mathcal{P}$ materializes at time $t_{*}$  \textit{cannot be predicted with certainty} -- Quantum Mechanics only enables us to calculate the ``frequency'' or probability by which a specific element $\Pi_{\xi} \in \mathcal{P}$ corresponds to the event materializing around time $t_{*}$, and this probability is given by \textit{Born's Rule}.
In colloquial language, one may say that if one knows the state in which an isolated physical system $S$ was prepared before the first event occurs then one can predict (using quantum mechanics) ``which pointer (of an instrument) will start to turn first, at approximately what time it will start to turn, and for how long it will turn before it will come to rest; but its final position cannot be predicted.''

(v) Note that many or most quantum-mechanical models of isolated systems that we discuss in our courses and books, such as models of systems of finitely many oscillators or of atoms treated according to Schr\"{o}dinger's wave mechanics and \textit{not} coupled to the quantized radiation field, \textit{do not describe any events} (in the sense this notion has been given above)! The reason is that they give rise to algebras $\mathcal{E}_{\geq t}$ that are \textit{independent} of $t$; i.e., that they do not exhibit any ``loss of access to information'', in the sense of Eq. \eqref{infoloss}. Before one incorporates equipment (with infinitely many degrees of freedom), such as detectors, etc., which the degrees of freedom of interest (e.g., the ones describing an atom) interact with, into the quantum-mechanical description it is \textit{impossible} to formulate a logically coherent theory of events and observations.
 
 Furthermore, one must expect that most systems have states, called  ``passive states'', with the property that there won't be any events happening \textit{even if} there is ``loss of access to information'', in the sense of Eq. \eqref{infoloss}. The reason is that the centers $\mathcal{Z}_{\rho_t}$ of the centralizers $\mathcal{C}_{\rho_t}$ of the algebras 
 $\mathcal{E}_{\geq t}$ may turn out to be trivial, for all times $t$, (or be independent of $t$), for \textit{certain} states $\rho$ (called ``passive''). One may even expect that, generically, a state is passive, and that equilibrium states at positive temperature are passive states. 

(vi) It is conceivable that, in a more elaborate formulation of quantum mechanics, there is no 
need to specify the list $\mathcal{O}_{S}$ of physical quantities of an isolated physical system $S$ that can, in principle, be detected directly. Rather, one can imagine that the algebras $$\lbrace \mathcal{Z}_{\rho_t} \vert t \in \mathbb{R}, \rho \text{ an arbitrary state on  } \mathcal{E} \text{  of physical interest}\rbrace$$ will \textit{determine} $\mathcal{O}_{S}$.

\subsection{The effective time evolution of states of systems featuring events}
\label{ETH}
 
Equations \eqref{Born} and \eqref{collapse} clarify the nature of the time evolution of states of systems featuring events. It is illustrated in the following Figure 3, where:

{\color{red}$E$} \text{   }stands for ``event'' (meaning that an event corresponding to a projection $\Pi_{\xi}$

\hspace{0.4cm} from some family $\mathcal{P}$ belonging to a maximal abelian subalgebra $\mathcal{M}_{S}^{(i)} \in \mathcal{O}_{S}$, 

\hspace{0.4cm} $i \in I_{S}$, materializes)

{\color{green}$T$ }  stands for ``tree'' (of states of $S$ corresponding to observed events, according

\hspace{0.4cm} to Eq. \eqref{collapse}); and

{\color{blue}$H$ }  stands for ``history'' (of observed events)

\vspace{0.2cm}
\noindent We thus speak of the \textit{``ETH approach''} to the interpretation of quantum mechanics (describing the effective quantum-mechanical time evolution of states of systems that feature events).
\begin{center}
\includegraphics[width=7.8cm, height=6cm]{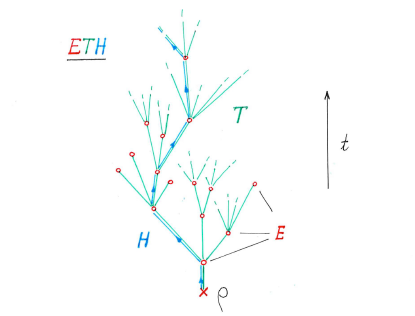}
\end{center}
\begin{center}
Figure 3: ``ETH approach'' to Quantum Mechanics
\end{center}
\noindent Let us summarize some basic elements of the ``ETH approach''. 
\begin{enumerate}
\item{
\textit{``Observables at infinity'':} 
Under rather general hypotheses, one can show that the algebra
$$\mathcal{E}_{\infty}:= \bigcap_{t\in \mathbb{R}} \mathcal{E}_{\geq t}$$
is \textit{abelian}, and that it is in the center of all the algebras $\mathcal{E}_{\geq t}, t \in \mathbb{R}$. Hence 
$\mathcal{E}_{\infty}$ is contained in the centers $\mathcal{Z}_{\rho_t}$ of the centralizers of all states $\rho_t = \rho\vert_{\mathcal{E}_{\geq t}}$, for an arbitrary state $\rho$ on $\mathcal{E}$ and all times $t$. Thus, the states $\rho_t$ can be decomposed over the spectrum, 
$\mathcal{X}_{\infty}$, of the algebra $\mathcal{E}_{\infty}$. Points in $\mathcal{X}_{\infty}$ are called \textit{``facts''}, because they correspond to objective values of \textit{time-independent} physical properties represented by operators in $\mathcal{E}_{\infty}$; see also \cite{BH}.
}
\item{
In \cite{Griff} and \cite{G-MH} the notion of ``consistent histories'' has been introduced and discussed, which, in conjunction with some understanding of the phenomenon of ``decoherence'' (see Sect. 4), is supposed to lead to a logically coherent interpretation of quantum mechanics. The problems with the approach in \cite{Griff, G-MH} are: (1) that there tend to exist many ``consistent histories'' that are \textit{incompatible} with each other, hence mutually exclude one another; and (2) that, in the understanding of the theory presented in these papers, the propagator of a system and the choice of an initial state do \textit{not} determine \textit{which} physical quantities that give rise to consistent histories will actually be observed in the course of time. Given the time evolution of a system and its initial state, the choice of a sequence of physical quantities giving rise to consistent histories thus remains quite arbitrary, i.e., lies -- like beauty -- ``in the eye of the beholder''. \\
This problem is avoided in the ``ETH approach'', as we now briefly explain.
Suppose that, at some time $t_0$, a physical system $S$ is prepared in a state $\rho=\rho^{0}$ (on the algebra $\mathcal{E}_{\geq t_0}$). Our formalism then enables us to \textit{predict} a time, $t_1$, around which the first event after the preparation of $S$ in state $\rho_0$ materializes and a family, 
$\mathcal{P}^{1} \subset \mathcal{O}_{S}$, of possible events, $\Pi_{\xi}^{1} \in \mathcal{P}^{1}$, to which the event that materializes at time $t_1$ belongs; see Eqs. \eqref{mixtstate} and \eqref{distance}. Suppose now that the event happening at time $t_1$ corresponds to the projection $\Pi_{\xi_1}^{1} \in \mathcal{P}^{1}$. The ``fundamental axiom'' (see item 2, \eqref{collapse}) then instructs us that, in order to improve our predictions of  the future after time $t_1$, we should use the state
$$\rho^{1}(\cdot):= \rho^{0}(\Pi_{\xi_1}^{1}(t_1))^{-1} \rho^{0}(\Pi_{\xi_1}^{1}(t_1)(\cdot)\Pi_{\xi_1}^{1}(t_1)),$$
on the algebra $\mathcal{E}_{\geq t_1}$, where $\Pi_{\xi_1}^{1}(t_1)$, (with $\Pi_{\xi_1} \in \mathcal{P}^{1}$), is the orthogonal projection on $\mathcal{H}$ describing the event happening at time $t_1$. Given the state $\rho^{1}$, one can now \textit{predict} a time $t_2$ and a family $\mathcal{P}^{2} \subset \mathcal{O}_{S}$ of orthogonal projections with the property that an event corresponding to some element of $\mathcal{P}^{2}$ happens around time $t_2$; etc.

Suppose that the state of the system after the $k^{th}$ event has happened around time $t_k$ is given by 
$\rho^{k}$, which is a state on the algebra $\mathcal{E}_{\geq t_k}, k=1,2,3,...$ . This state determines a time $t_{k+1} > t_k$ and a family, $\mathcal{P}^{k+1}$, of orthogonal projections describing possible events that might materialize at time $t_{k+1}$. (It can happen that $\rho^{k}$ is a ``passive state'', in which case 
$t_{k+1} = \infty $.) Suppose that the event happening at time $t_{k+1}$ is detected to be given by the operator $\Pi_{\xi_{k+1}}^{k+1}(t_{k+1})$ representing the projection 
$\Pi_{\xi_{k+1}}^{k+1} \in \mathcal{P}^{k+1}$. According to Eq. \eqref{collapse}, the state on the algebra 
$\mathcal{E}_{\geq t_{k+1}}$ to be used to predict the future at times $>t_{k+1}$ is then given by
\begin{equation}
\label{rec-k+1}
\rho^{k+1}(\cdot):= \rho^{k}(\Pi_{\xi_{k+1}}^{k+1}(t_{k+1}))^{-1} \rho^{k}(\Pi_{\xi_{k+1}}^{k+1}(t_{k+1})(\cdot)\Pi_{\xi_{k+1}}(t_{k+1})).
\end{equation}
If, however, the event happening at time $t_{k+1}$ (representing some element of $\mathcal{P}^{k+1}$) is \textit{not} recorded then the state on $\mathcal{E}_{\geq t_{k+1}}$ to be used to predict the future after time $t_{k+1}$ is given by
\begin{equation}
\label{notrec}
\rho^{k+1}(\cdot):= \rho^{k}(\cdot)\vert_{\mathcal{E}_{\geq t_{k+1}}} \simeq \sum_{ \Pi_{\xi}^{k+1} \in \mathcal{P}^{k+1}} \rho^{k} (\Pi_{\xi}^{k+1}(t_{k+1})(\cdot) \Pi_{\xi}^{k+1}(t_{k+1})).
\end{equation}
Recall that the distance between $\mathcal{P}^{k+1}(t_{k+1})$ and 
$\mathcal{Z}_{\rho^{k}_{t_{k+1}}} (\subset \mathcal{E}_{\geq t_{k+1}})$ is tiny!

\vspace{0.2cm}
 In the ``ETH approach'', a \textit{history} consists of a sequence, $\big(t_k, \Pi_{\xi_k}^{k}(t_k)\big)_{k=1, 2, 3, ...}$, where
 $t_1 < t_2 < t_3 < ...$ are times, $\Pi_{\xi_k}^{k}(t_k)$ (with $\Pi_{\xi_k}^{k} \in \mathcal{P}^{k} \subseteq \mathcal{O}_{S}$) 
is the orthogonal projection on $\mathcal{H}$ describing the event happening at approximately time $t_k$, with $t_k$ and $\mathcal{P}^{k}$ determined by the state $\rho^{k-1}$ corresponding to the event that happened at time $t_{k-1}$, according to Eq. \eqref{collapse}.

 Such a history is denoted for short by 
 \begin{equation}
 \label{hist}
 \lbrace (\xi_k, t_k)\vert k=1,2,3,... \rbrace
 \end{equation} 
 Events that have materialized at some time, but have not been recorded can    
 be omitted from the list \eqref{hist} -- as follows from Eq. \eqref{notrec}.\\
Quantum mechanics, as understood in the ``ETH approach'', predicts the probabilities of histories. In fact, 
these probabilities are given by a well-known formula, which we call \textit{``LSW-formula''} 
(for ``L\"{u}ders-Schwinger-Wigner'', see \cite{Lu, Sch, Wig}). It is the unique generalization of Born's Rule 
($k=1$) to all values of $k$. Here it is:
\begin{equation}
\label{LSW}
\text{Prob}\lbrace((\xi_k, t_k)\vert k=1,2,3,... \rbrace := \rho^{0}\Big(\prod_{k=1, 2, 3, ...} \Pi_{\xi_k}^{k}(t_k)\cdot 
(\prod_{k=1, 2, 3, ...}\Pi_{\xi_k}^{k}(t_k))^{*} \Big)
\end{equation}
Some applications of this formula will be sketched in Sect. 3.
}
\item{
It should be emphasized that a physical quantity represented by an operator $\hat{X} \in \mathcal{O}_{S}$ that, for a suitably chosen initial state, has an objective value around some time $t$ -- meaning that the spectral projections of $X(t)$ belong to a family $\mathcal{P}(t)$ of possible events happening at time $t$ -- will usually \textit{not} have an objective value at an earlier or later time, because the quantity in question evolves in time; i.e., the operators $X(t)$ representing that quantity depend on time $t$. In fact, for typical choices of an element $\hat{X} \in \mathcal{O}_{S}$, the operators $X(t)$ do \textit{not} commute with operators describing ``conserved quantities'', such as energy, momentum or angular momentum, etc. (They \textit{do} however commute with operators representing ``Super-Selection Rules''. But energy and momentum are, of course, not Super-Selection Rules.) It then follows that the detection of the value of the physical quantity represented by an operator $\hat{X} \in \mathcal{O}_{S}$, with the property that $X(t)$ depends on $t$, \textit{violates} energy- (and possibly angular momentum-  \dots) conservation, in the sense that the distribution of energies (and angular momentum, i.e., the energy- and angular-momentum fluctuations) in the states before and after the observation of the value of $\hat{X}$, see Eq. \eqref{collapse}, are \textit{different} from each other. \\
The relation between the \textit{duration} of the event corresponding to the recording of the value of a physical quantity and the \textit{amount of energy fluctuation} accompanying this event is given by the usual \textit{time-energy uncertainty relations}; see, e.g., \cite{PfF}.
}
\end{enumerate}

\section{Indirect observation/reconstruction of properties of physical systems}
\label{Kraus-meas}

In this section, we present a brief outline of the theory of indirect non-demolition measurements, as originally developed in \cite{Kraus}; see also \cite{MK, BB, BFFS} and references given there. Our discussion is limited to the analysis of a simple example, which is inspired, in part, by the beautiful experiments described in \cite{Haroche}.

\subsection{The example of a mesoscopic system}

A \textit{property}, $P$, of a physical system $S$ is the value of a \textit{time-independent} physical quantity. Examples of properties of $S$ that can be recorded directly are ``observables at infinity'', as described in item 1 of subsect. \ref{ETH}. Our purpose, in this section, is to present a sketchy outline of how properties of a system $S$ can be determined \textit{indirectly} from recordings of long sequences of events, (i.e., from recordings of the values of physical quantities represented by operators belonging to $\mathcal{O}_{S}$), as discussed in the last section. Such an indirect observation of a property $P$ of $S$ is sometimes called a \textit{``non-demolition measurement''}. A presentation of the general theory of non-demolition measurements is beyond the scope of this paper; but see \cite{Kraus, MK, BB, BFFS}. In the following, we therefore focus our attention on the concrete example of a mesoscopic system $S$ sketched at the end of subsection \ref{iloss}. In this example, the list $\mathcal{O}_{S}$ of physical quantities whose values can be recorded directly consists of all bounded functions of a single operator, namely the operator $\hat{X}$ defined in Eq. \eqref{X} of subsect. \ref{iloss}, which has only two eigenvalues $ \pm 1$ corresponding to projections $\Pi_{\pm 1}$.

We imagine that a history, $(\xi_k, t_k)_{k=1, 2, 3, ...}$, of events corresponding to values \mbox{$\xi_{k} = \pm 1$} of the physical quantity  represented by the operator $\hat{X}$ has been recorded. In our example, the recording of the value $\xi_k=1$ at time $t_k$ means that the detector near $D_L$ has clicked around time $t_k$, (i.e., an electron traveling through the T-channel has entered the reservoir at the end of the left arm of the T-channel), while the recording of $\xi_k=-1$ at time $t_k$ means that the detector near $D_R$ has clicked around time $t_k$.

For the following discussion, the values of the times $t_1 < t_2 < t_3 < ...$ at which events (i.e., clicks of a detector) are happening are unimportant. We therefore omit reference to these times in our notations, denoting histories by  $\underline{\xi}=(\xi_k)_{k=1, 2, 3, ...}$, with $\xi_{k}=\pm 1, \forall k=1,2,3,...$. By $\Xi$ we denote the space of all arbitrarily long histories. A sequence, $\underline{\xi}^{n}:=(\xi_k)_{k=1}^{n}$, of $n$ recorded detector clicks belonging to a history $\underline{\xi}$ is called a \textit{``measurement protocol''} of length $n$.
 
Given an initial state $\rho$ of $S$, the ``LSW formula'', Eq. \eqref{LSW} of subject. \ref{ETH}, determines a probability measure $\mu_{\rho}$ on the space $\Xi$: The probability of a measurement protocol 
$\underline{\xi}^{n}=(\xi_1, ..., \xi_n)$ of length $n$ is given by 
\begin{equation}
\label{LSW1}
\mu_{\rho}(\xi_1, ..., \xi_{n}):= \rho\Big(\Pi_{\xi_1}(t_1)\cdots \Pi_{\xi_n}(t_n) \cdots \Pi_{\xi_1}(t_1) \Big),
\end{equation}
 see \eqref{LSW}. We note that 
 $$\sum_{\xi_n} \mu_{\rho}(\xi_1,...,\xi_{n-1}, \xi_n) = \mu_{\rho}(\xi_1,..., \xi_{n-1}), \quad \text{   and }\quad \sum_{\underline{\xi}^{n}} \mu_{\rho}(\underline{\xi}^{n}) = 1.$$
 
By a lemma due to Kolmogorov, these properties imply that $\mu_{\rho}$, as defined by \eqref{LSW1}, extends to a probability measure on the space $\Xi$ of histories.
 
 We suppose that the chemical potential of the reservoir ``$e^{-}$ gun'' is only very slightly higher than the chemical potentials of the reservoirs $D_L$ and $D_R$, so that the rate, $\tau$, at which ``$e^{-}$ gun'' releases an electron into the T-channel is so slow that, at any given moment, there is typically only at most one electron traveling through the T-channel, and that, after an electron has entered $D_{\ell}$, the state of this reservoir and of the detector near it relaxes to the original state in a time much shorter than $\tau$, for $\ell = L, R$. These assumptions can be interpreted as saying that the electrons traveling through the T-channel -- to get lost in one of the reservoirs, $D_L$ or $D_R$, at the end of the horizontal arms of the T-channel -- and their successive detections are all \textit{independent} of each other. This implies that the measures 
 $\mu_{\rho}$ are \textit{``exchangeable''}, i.e., 
 \begin{equation}
 \label{exch}
 \mu_{\rho}(\xi_1,...,\xi_n)=\mu_{\rho}(\xi_{\sigma(1)},...,\xi_{\sigma(n)}), \quad \forall \text{   permutations   } \sigma\text{  of   }\lbrace 1,...,n \rbrace,
 \end{equation}
for all $n=1, 2, 3, ...$ and all states $\rho$ of the system whose restriction to the three reservoirs have the desired properties, (in particular, the prescribed chemical potentials).

By de Finetti's theorem, Eq. \eqref{exch} implies that $\mu_{\rho}$ is a convex combination of \textit{product measures}. For simplicity, we suppose that it is a \textit{finite} convex combination of product measures:
\begin{equation}
\label{deFin}
\mu_{\rho}(\xi_1,...,\xi_n)=\sum_{\nu=0}^{N} \pi_{\rho}(\nu) \prod_{i=1}^{n} p(\xi_i \vert \nu),
\end{equation}
where
$$p(\xi \vert \nu) \geq 0, \quad \forall \text{  } \xi, \nu, \text{   }\text{   and    } \sum_{\xi=\pm 1} p(\xi \vert \nu) =1, \quad \forall \text{   } \nu=0, 1, ...,N,$$
and 
$$ 0 \leq \pi_{\rho}(\nu) <1, \quad \forall \text{   }\nu, \quad \text{with    } \sum_{\nu=0}^{N} \pi_{\rho}(\nu) =1.$$
The physical interpretation of these quantities is as follows:
\begin{itemize}
\item{$\nu$ is the number of electrons bound by the quantum dot $P$. Because of possible electron exchange between $P$ and $P'$, the state $\rho$ of $S$ is, in general, \textit{not} an eigenstate of the electron number operator of $P$; i.e., $\nu$ does usually \textit{not} have a sharp value in the state $\rho$. It is assumed, however, that $\nu$ is a \textit{static} quantity, i.e., that the electron number operator of $P$ commutes with the Hamiltonian of the system.}
\item{$p(\xi \vert \nu)$ is the a-priori probability that an electron traveling through the T-channel reaches the detector near $D_L$ ($\leftrightarrow \xi = 1$) or the one near $D_R$ ($\leftrightarrow \xi = -1$), respectively. This probability clearly depends on the number $\nu$ of electrons bound to the dot $P$, because these electrons create a ``Coulomb blockade'' in the arm of the T-channel above $P$ and extending to the right, towards $D_R$.}
\item{$\pi_{\rho}(\nu)$ is the Born probability (in the state $\rho$) for the number of electrons bound to the dot $P$ to be equal to $\nu$, with $\nu=0,...,N$.}
\end{itemize}

\subsection{Summary of results on indirect measurements}

In this last subsection, we summarize some recent results on the system described above. We omit the proofs, which the reader may find in \cite{BFFS}.

We define the frequency, $f_{\xi}^{(n)}$, of the value $\xi$ in a measurement protocol $\underline{\xi}^{n}$ of length $n$ (with $\xi=1 \leftrightarrow$ electron reaches $D_L$, $\xi=-1 \leftrightarrow$ electron reaches $D_R$) as follows:
\begin{equation}
\label{frequ}
f_{\xi}^{(n)}(\underline{\xi}):= \frac{1}{n} \Big(\sum_{k=1}^{n} \delta_{\xi\text{  }\xi_k} \Big), \quad \text{  with    } \text{   }
\sum_{\xi= \pm 1} f_{\xi}^{(n)}(\underline{\xi}) = 1, \quad \forall n.
\end{equation}

The following results have been established in \cite{BFFS}.
\vspace{0.4cm}

1. \underline{Law of Large Numbers}\\

\hspace{0.5cm}For every history $\underline{\xi}$,
\begin{equation}
\label{LoLN}
\underset{n \rightarrow \infty}{\text{lim}  } f_{\xi}^{(n)}(\underline{\xi}) = p(\xi \vert \nu),
\end{equation}

\hspace{0.5cm} for some $\nu=0, 1, ..., N$. $\Box$\\

For simplicity, we assume that
\begin{equation}
\label{gap}
\underset{\nu_1 \not= \nu_2}{\text{min}} \vert p(1\vert \nu_1) - p(1\vert \nu_2)\vert \geq \kappa > 0.
\end{equation}
With each $\nu = 0,1,.., N$ we associate a subset, $\Xi_{\nu}$, of $\Xi$ defined by
\begin{equation}
\label{subset}
\Xi_{\nu}(n, \underline{\varepsilon}):=\lbrace \underline{\xi} \in \Xi \vert \vert f_{\xi}^{(n)}(\underline{\xi}) - 
p(\xi \vert \nu) \vert < \varepsilon_{n} \rbrace,
\end{equation}
where
$$\varepsilon_{n} \rightarrow 0, \sqrt{n} \varepsilon_{n} \rightarrow \infty, \quad \text{   as   } n\rightarrow \infty.$$

2. \underline{Disjointness}\\

\hspace{0.4cm} It follows from assumption \eqref{gap} and definition \eqref{subset} that, for $n$ so large that

\hspace{0.4cm} $\varepsilon_{n} < \frac{\kappa}{2}$,
\begin{equation}
\label{disjointness}
\Xi_{\nu_1}(n, \underline{\varepsilon}) \cap \Xi_{\nu_2}(n, \underline{\varepsilon}) = \emptyset, \quad \nu_1 \not= \nu_2. \qquad \Box
\end{equation}

3. \underline{Born's Rule and Central limit Theorem}\\

\hspace{0.4cm} Under appropriate hypotheses on the state $\rho$ (see \cite{BB,BFFS}),
\begin{equation}
\underset{n \rightarrow \infty}{\text{lim}  } \mu_{\rho}(\Xi_{\nu}(n, \underline{\varepsilon})) = \pi_{\rho}(\nu)
\end{equation}
\begin{center}
\textit{     Born's Rule}
\end{center}

\hspace{0.4cm} Furthermore,
\begin{equation}
\label{full-measure}
\mu_{\rho} \Big( \bigcup_{\nu} \Xi_{\nu}(n, \underline{\varepsilon}) \Big) \rightarrow 1, \quad \text{  as    } n \rightarrow \infty. \qquad \Box
\end{equation}\\

\vspace{2cm}

4. \underline{Theorem of Boltzmann-Sanov}\\

\hspace{0.4cm} Defining the relative entropy $\sigma(\nu_1\Vert \nu_2)$ by
$$\sigma(\nu_1 \Vert \nu_2):= \sum_{\xi = \pm 1} p(\xi\vert \nu_{1}) \big( \text{log}_{2} \text{  }p(\xi\vert \nu_{1}) - 
\text{log}_{2} \text{  }p(\xi\vert \nu_{2}) \big),$$

\hspace{0.4cm} one has that 
\begin{equation}
\label{BoltzSan}
\mu \big(\Xi_{\nu_1}(n, \underline{\varepsilon})\vert \nu_2 \big) \leq C\text{   }e^{-n\sigma(\nu_1 \Vert \nu_2)},
\end{equation}

\hspace{0.4cm} where $\mu(\cdot \vert \nu)$ is the product measure determined by $p(\xi \vert \nu)$.\\

\textit{Remark:} Results 1 through 3 hold in much greater generality; see \cite{BFFS}. Concerning Eq. \eqref{deFin} and Result 4, we remark that  there is a general theory of how to decompose measures $\mu_{\rho}$ into ``extremal measures'', $\mu(\cdot \vert \nu), \nu \in \Xi_{\infty}$, 
where $\Xi_{\infty}$ is the spectrum of the algebra of functions on $\Xi$ measurable at $\infty$. 
(Functions on $\Xi$ measurable at $\infty$ take the same values on any two histories
$\underline{\xi}$ and $\underline{\eta}$ with $\xi_k = \eta_k$, except for finitely many $k$.) One can show that, under suitable assumptions, extremal measures are determined again by states of the system via the ``LSW formula''.

We pause to interpret Results 1 through 4. It follows from \eqref{full-measure} that if $n$ is very large then the set $ \bigcup_{\nu} \Xi_{\nu}(n, \underline{\varepsilon})$ has apparently nearly full measure with respect to 
$\mu_{\rho}$. By \eqref{disjointness}, it then follows that, for very large $n$, essentially every history
 $\underline{\xi}$ belongs to exactly one of the sets 
$\Xi_{\nu}(n, \underline{\varepsilon})$, and hence a measurement protocol $\underline{\xi}^{n}$ of length $n$ determines the number $\nu$ of electrons bound to the dot $P$ nearly unambiguously, with an error margin that tends to $0$, as $n$ tends to $\infty$. In the limit 
$n \rightarrow \infty$, measurement protocols determine the number $\nu$ of electrons in the dot $P$ precisely, which implies that this number becomes sharp (i.e., does not exhibit any fluctuations, anymore), as $n$ tends to $\infty$. This is the phenomenon of \textit{``purification''} first studied in \cite{MK}.
Furthermore, the empirical probability of a history $\underline{\xi} \in \Xi_{\nu}(n, \underline{\varepsilon})$ tends to $\pi(\nu)$, as $n$ tends to $\infty$, which establishes Born's Rule for non-demolition measurements.

Finally, by Result 4 (Boltzmann-Sanov), the \textit{time}, $T$, it takes to indirectly determine the number $\nu$ of electrons
bound to the dot $P$ is given, approximately, by
\begin{equation}
\label{detectiontime}
T=\tau/\sigma,
\end{equation}
where $\tau$ is the rate at which ``$e^{-}$ gun'' shoots electrons into the T-channel (i.e., the time elapsing between two consecutive electrons traveling through the T-channel, or two consecutive clicks of detectors), and 
$$\sigma := \underset{\nu_1 \not= \nu_2}{\text{min}} \sigma(\nu_1 \Vert \nu_2).$$

It should be emphasized that most indirect measurements are \textit{not} non-demolition measurements. In the example of the mesoscopic system studied above, it is an idealization to assume that the number of electrons in the dot $P$ is static (i.e., that the electron number operator counting the number of electrons bound to $P$ commutes with the Hamiltonian of the system). It is therefore important to generalize the theory of indirect measurements sketched here to situations where properties of a system $S$ change in time, albeit much more slowly than the rate at which direct observations of physical quantities in 
$\mathcal{O}_{S}$ are made. A beginning of such a theory has been described in \cite{BFFS}.\\

\section{Some hints to the literature, conclusions}

There are many precursors of some of the ideas described in this paper, and it is quite impossible to do justice to all authors who have contributed (more and less) important pieces to the mosaic. The puzzling features of quantum mechanics and the problems surrounding its interpretations have been discussed by many people, including Schr\"{o}dinger \cite{Schrod} and, later on, Bell; see \cite{Bell}. A general reference where many of the (older) interpretations of quantum mechanics are described is \cite{WZ}. \footnote{The Bohmian point of view, which is of definite interest, but is not relevant for the material in this paper, is presented in detail in \cite{DT}, and references given there. For the many-worlds interpretation of quantum mechanics, see \cite{Everett, Dewitt}.} By no means are we attempting to distribute credit to various schools of thought, and we offer our apologies to all those colleagues who feel that their work should be cited here, but isn't. However, in all modesty, we feel that we have developed a novel approach to a ``quantum theory of events and experiments'', and we hope that the reader may have profited from reading this summary of some of the key ideas underlying our approach. (More details concerning the approach summarized in this paper can be found in \cite{FFS, FS2}.)

One might say that ``loss of access to information'', in the sense of Eq. \eqref{infoloss}, is a special form of what is called \textit{``decoherence''}. The concept of decoherence was introduced and discussed in \cite{Ludwig, Zeh, Primas} and, in a very clear way, in \cite{Hepp} and further developed in \cite{Zurek, Blanchard}, and references given there. There cannot be any doubt that ``decoherence'' is a basic building block in a quantum theory of events and experiments. Attempts to arrive at a logically coherent theory of observations and measurements based on the concepts of ``consistent histories'' and ``decoherence'' have been presented in \cite{Griff, G-MH, Omnes}. \footnote{We refrain from discussing the merits and success of various attempts to interpret quantum mechanics -- which does not mean that we do not have opinions about them.}
The crucial concept of an ``event'' was introduced and discussed, and its importance in understanding the deeper meaning of quantum mechanics emphasized, in \cite{Haag}. This thread of thoughts has been taken up in \cite{BJ}, \cite{Jadc}, where a formulation of an ``event-enhanced quantum theory'' inspired by \cite{Haag} and \cite{GRW} has been proposed. 

In our approach, the concept of an ``event'' is given a clear meaning, and it plays a fundamental role; see 
subsects. \ref{events} and \ref{ETH}. Our formulation of the quantum theory of systems exhibiting ``loss of access to information'', in the sense of Eq. \eqref{infoloss}, subsect. \ref{iloss}, and ``events'', as defined in the ``fundamental axiom'' of subsect. \ref{events}, introduces a clear distinction between the past and the future: the past is factual -- it consists of events that have materialized --, while the future consists of potentialities, namely of ``possible events'' that might happen, but need not happen.

When analyzing a problem such as the deeper meaning of Quantum Mechanics one must fear that not all readers will find one's approach to the problem entirely convincing. Most certainly, our analysis is no exception. Moreover, we realize that various rather interesting and important technical issues concerning our approach remain open; (although there is no reason why one should not be able to settle them).
We therefore conclude our report with a famous quote:\\

\textit{``Wir stehen selbst entt\"{a}uscht und sehn betroffen}\\
\indent{\textit{den Vorhang zu und alle Fragen offen. ... }}\\

\textit{Verehrtes Publikum, los, such dir selbst den Schluss!}\\
\indent{\textit{Es muss ein guter da sein, muss, muss, muss!''}}\\

(Bertolt Brecht, in: \textit{``Der gute Mensch von Sezuan''})\\

This paper is a token of our deep gratitude for the wisdom Raymond Stora has shared with us and the friendship he has bestowed upon us.

\vspace{0.5cm}
\textit{Acknowledgements.}

We thank our many colleagues, collaborators and friends -- too many to mention them all explicitly -- for all they have contributed to improving our understanding of quantum mechanics. Ph. B. is grateful for many illuminating discussions with the late R. Haag, and he thanks M. Hellmich, A. Jadczyk  and R. Olkiewicz for fruitful collaborations. J. F. and B. S. thank M. Ballesteros and M. Fraas for numerous exchanges and collaboration.\\
J. F. thanks P.-F. Rodriguez for discussions and for drawing his attention to the short story of J. L. Borges that we have quoted from, and D. Brydges for useful comments on a first draft of this paper. This paper was written while J. F. was visiting the School of Mathematics of The Institute for Advanced Study at Princeton. He acknowledges financial support from The Giorgio and Elena Petronio Fellowship Fund, and he thanks Tom Spencer for his generous hospitality.\\

\vspace{1cm}
\appendix
{\Large\bf{Appendix: Proofs of Eqs. \eqref{centralizer} and \eqref{center}}}\\

We first briefly explain the notion of a ``centralizer'', $\mathcal{C}_{\varphi}$, of a state, $\varphi$, on a von Neumann algebra, $\mathcal{M}$. We recall that
$$\mathcal{C}_{\varphi} := \lbrace X \in \mathcal{M} \vert \text{ad}_{X}(\varphi)(\cdot):=\varphi([X,\cdot])=0 \rbrace.$$
It follows from this definition that $\mathcal{C}_{\varphi}$ is a \textit{subalgebra} of $\mathcal{M}$: If $X$ and $Y$ are elements of $\mathcal{C}_{\varphi}$ then, obviously, any linear combination of $X$ and $Y$ belongs to 
$\mathcal{C}_{\varphi}$, too. Furthermore, for arbitrary $A \in \mathcal{M}$,
$$\varphi\big(XYA\big)=\varphi\big((YA)X\big)= \varphi\big(Y(AX) \big) =\varphi\big((AX)Y\big)
=\varphi \big(AX Y \big),$$
i.e., $XY$ belongs to $\mathcal{C}_{\varphi}$, too.\\
Next, let $X=X^{*}$ belong to $\mathcal{C}_{\varphi}$, with
$$X=\sum_{j=1}^{N}\xi_{j}\Pi_{\xi_j}$$ the spectral decomposition of $X$, where $\xi_1< \cdots <\xi_N$ are the eigenvalues of $X$ and $\Pi_{\xi_1},\dots,\Pi_{\xi_N}$ its spectral projections. Since any polynomial in 
$X$ belongs to $\mathcal{C}_{\varphi}$, too, it follows that $\Pi_{\xi_j} \in \mathcal{C}_{\varphi}$, for any $j=1,...,N$.
Thus, for an arbitrary operator $A \in \mathcal{M}$,
\begin{eqnarray}
\label{incoherent}
\varphi(A) & = & \sum_{i,j=1,...,N} \varphi \big(\Pi_{\xi_i}A \Pi_{\xi_j}\big) \nonumber\\
                   & = & \sum_{i,j=1,...,N} \varphi \big(A \Pi_{\xi_j} \delta_{ij} \big) \nonumber\\
                   & = & \sum_{i=1}^{N} \varphi \big(\Pi_{\xi_i} A \Pi_{\xi_i} \big)
\end{eqnarray}
Conversely, if Eq. \eqref{incoherent} holds for arbitrary $A\in \mathcal{M}$ then it obviously follows that $X$ belongs to $\mathcal{C}_{\varphi}$.
It is obvious that Eq. \eqref{incoherent} also holds if $X$ belongs to the center, $\mathcal{Z}_{\varphi}$, of 
$\mathcal{C}_{\varphi}$.\\

\textit{Application: Proofs of Eqs. \eqref{centralizer} and \eqref{center}.}\\

Let $X(t)=\sum_{j=1}^{N} \xi_{j} \Pi_{\xi_j}(t)$ be as in \eqref{spect-thm}, and let $A$ be an arbitrary operator in $\mathcal{E}_{\geq t}$. We rewrite $\rho_{t}(A)$ as follows:

\begin{eqnarray}
\rho_{t}(A) & = & \sum_{i,j = 1}^{N}\lbrace [\rho_{t}\big(\Pi_{\xi_i}(t)A\Pi_{\xi_j}(t)\big) - \rho_{t}    
                       \big(E_{\rho_t}(\Pi_{\xi_i}(t))A\Pi_{\xi_j}(t)\big) ] \nonumber\\
                      &  & +[\rho_{t}\big(A\Pi_{\xi_j}(t)E_{\rho_t}(\Pi_{\xi_i}(t))\big) -\rho_{t}\big(A\Pi_{\xi_j}(t)\Pi_{\xi_i}(t))    
                      \big) ]\rbrace \nonumber\\
                      &  & + \sum_{i=1}^{N} \lbrace \rho_{t}\big(A\Pi_{\xi_i}(t)^{2}\big) - \rho_{t}\big(A\Pi_{\xi_i}
                      (t)E_{\rho_t}(\Pi_{\xi_i}(t))\big)] \nonumber\\
                      &  & + [\rho_{t}\big(E_{\rho_t}(\Pi_{\xi_i}(t))A \Pi_{\xi_i}(t)\big) - \rho_{t}\big(\Pi_{\xi_i}(t)A \Pi_{\xi_i}
                      (t)\big)] \rbrace \nonumber\\
                      &  &  + \sum_{i=1}^{N}\ \rho_{t}\big(\Pi_{\xi_i}A\Pi_{\xi_i}(t)\big)
 \end{eqnarray}

In the second and in the fourth line we have used Eq. \eqref{defcentral}, which is legitimate, because $E_{\rho_t}(\Pi_{\xi_i}(t))$ belongs to the centralizer $\mathcal{C}_{\rho_t}$ of $\rho_t$. Since we are assuming that 
$\Vert E_{\rho_t}(\Pi_{\xi_j}(t)) - \Pi_{\xi_j}(t) \Vert \leq \delta', \text{  } \forall j=1,...,N,$ it follows that the absolute values of the four terms on the right side of line 1 and on lines 2, 3 and 4, respectively, are bounded above by $N\delta' \text{  } \Vert A \Vert$. This implies \eqref{mixtstate}, and hence \eqref{centralizer} is proven.

To prove \eqref{center}, all we have to do is to repeat the above argument with $E_{\rho_t}(\Pi_{\xi_i}(t))$ replaced by $e_{\rho_t}(\Pi_{\xi_i}(t))$, which belongs to $\mathcal{Z}_{\rho_t} \subseteq \mathcal{C}_{\rho_t}$.

\end{document}